\documentclass[a4paper,fleqn,usenatbib]{mnras}

\usepackage[T1]{fontenc}
\usepackage{ae,aecompl}

\usepackage{graphicx}	
\usepackage{amsmath}	
\usepackage{amssymb}	

\usepackage{ulem}
\usepackage[usenames]{color}
\definecolor{rred}{rgb}{0.8,0,0}

\title[LAT detection of transient gamma-ray emission]{\textit{Fermi}/LAT detection of a transient gamma-ray flare in the vicinity of the binary star DG CVn}
\author[A. Loh, S. Corbel \& G. Dubus]
{Alan Loh$^{1}$\thanks{E-mail: alan.loh@cea.fr},
St\'ephane Corbel$^{1, 2}$
and Guillaume Dubus$^{3, 4}$\\
$^{1}$Laboratoire AIM (CEA/IRFU - CNRS/INSU - Univ. Paris Diderot), CEA DSM/IRFU/SAp, F-91191 Gif-sur-Yvette, France\\
$^{2}$Station de Radioastronomie de Nan\c{c}ay, Observatoire de Paris, PSL Research University, CNRS, Univ. Orl\'{e}ans, 18330 Nan\c{c}ay, France \\
$^{3}$Univ. Grenoble Alpes, IPAG, F-38000 Grenoble, France\\
$^{4}$CNRS, IPAG, F-38000 Grenoble, France
}

\date{Accepted XXX. Received YYY; in original form ZZZ}

\pubyear{2016}

\begin{document}
\label{firstpage}
\pagerange{\pageref{firstpage}--\pageref{lastpage}}
\maketitle

\begin{abstract}
Solar flares are regularly detected by the Large Area Telescope (LAT) on board the \textit{Fermi} satellite, however no $\gamma$-ray emission from other stellar eruptions has ever been captured. The \textit{Swift} detection in April 2014 of a powerful outburst originating from DG\,CVn, with associated optical and radio emissions, enticed us to search for possible $0.1$--$100$\,GeV emission from this flaring nearby binary star using the {\it Fermi}/LAT. 
No $\gamma$-ray emission is detected from DG\,CVn in 2014, but we report a significant $\gamma$-ray excess in November 2012, at a position consistent with that of the binary. 
There are no reports of contemporary flaring at other wavelengths from DG\,CVn or any other source within the error circle of the $\gamma$-ray source. We argue that the $\gamma$-ray flare is more likely to have been associated with a background blazar than with DG\,CVn and identify a candidate for follow-up study.
\end{abstract}

\begin{keywords}
Acceleration of particles -- stars: flare -- gamma-rays: stars -- stars: individual (DG$\,$CVn).
\end{keywords}

\section{Introduction}
The development of wide-field surveys and rapid response capabilities at all wavelengths has enabled the discovery of unanticipated classes of transient sources \citep[see][for radio, optical and X-ray examples]{2011BASI...39..315F, 2009PASP..121.1334R, 2015JHEAp...7....2G}. 
For the high-energy sky above $100$\,MeV, the main instrument of the \textit{Fermi} satellite, the Large Area Telescope \citep[LAT,][]{2009ApJ...697.1071A}, combines a high sensitivity, a wide field of view, a large energy range, and operates in a sky-survey mode most of the time. 
This nearly complete mapping and continuous monitoring of the sky led to the discovery of new and sometimes unexpected $\gamma$-ray source classes such as microquasars \citep{2009Sci...326.1512F} or Galactic novae \citep{2010Sci...329..817A}. 

The hard X-ray transient monitor Burst Alert Telescope \citep[BAT,]{2005SSRv..120..143B} on board the \textit{Swift} satellite detected on 2014 April 23 a powerful and rare outburst \citep{2014ATel.6121....1D, 2016arXiv160904674O}. 
The brightness of this event was such that it triggered \textit{Swift} as if it were a Gamma-Ray Burst. 
The associated source of this emission, DG$\,$CVn (also known as GJ$\,$3789 or G$\,$165$-$8AB) is a stellar system comprised of two M-dwarf stars separated by $0\farcs2$ \citep[][]{2001AJ....122.3466M, 2004A&A...425..997B}. 
\citet{2014AJ....147...85R} indicated that the system lies at $18\,$pc from the Earth and that it is relatively young \citep[${\sim} 30\,$Myr, ][]{2015MNRAS.452.4195C}. 
Intense chromospheric activity in radio, H$\alpha$, and X-rays is associated with the rapid stellar rotation \citep[$v \sin i = 55.5\,$km$\,$s$^{-1}$,][]{1998A&A...331..581D, 2003ApJ...583..451M}.

\textit{Swift} team triggered an automatic follow-up with the Arcminute Microkelvin Imager radio telescope at $15\,$GHz reported by \citet{2015MNRAS.446L..66F}. 
Radio observations started within $6$ minutes after the trigger and captured a bright $100\,$mJy flare. 
Some additional smaller flares occurred during the next four days before the return at a quiescent radio level \citep[$2$--$3\,$mJy, as detected by][]{2009ApJ...701.1922B}.
DG$\,$CVn's radio detection suggests production of synchrotron emission from electrons accelerated during the initial phase of a major stellar flare. 
These non-thermal particles are thought to deposit their energy in the lower stellar atmosphere where the density is higher, heating the medium and possibly producing X-ray thermal radiation from the plasma \citep[e.g.][]{1968ApJ...153L..59N}. 
\citet{2015MNRAS.452.4195C} measured a delay between hard X-ray and optical emissions, that can be attributed to this Neupert effect.

The accelerated particles could also lose their energy via pion decay or Bremsstrahlung processes depending on their leptonic or hadronic nature. This may result in high-energy emission that could be detectable by the LAT.
This motivated the $\gamma$-ray study described in Sec.~\ref{sec:analysis} of this letter. 
Results and detection of a significant excess in 2012, close to DG\,CVn, are presented in Sec.~\ref{sec:results} and discussed in Sec.~\ref{sec:discussion}, where we consider the possibility that this excess is due to a flaring Active Galactic Nucleus (AGN).

\section{Fermi/LAT Data Analysis} \label{sec:analysis}
We have analysed the Pass 8 data gathered by the LAT since its launch in August 2008 until November 2015, seven years later. The reduction and analysis of the LAT products were performed using the 10-00-02 version of the \textit{Fermi} Science Tools\footnote{\href{http://fermi.gsfc.nasa.gov/ssc/data/analysis/documentation/}{http://fermi.gsfc.nasa.gov/ssc/data/analysis/documentation/}.} with the Instrument Response Functions set \texttt{P8R2\_SOURCE\_V6} \citep{2013arXiv1303.3514A}.

	\subsection{Analysis set-up}\label{sec:setup}
For the purpose of the $\gamma$-ray analysis, we have considered a $15\degr$ acceptance cone centred on DG$\,$CVn's position (at $\mbox{RA}=202\fdg94$, $\mbox{Dec.}=29\fdg28$, J2000). LAT photons labelled as \texttt{SOURCE} (\texttt{evclass=128}) inside this region were selected in the energy range from $100\,$MeV to $100\,$GeV. 
Furthermore, as the $\gamma$-ray excess near DG\,CVn's location appears to be soft ({i.e.,} most of the photon energies are below few GeVs, see \S\ref{sec:nov12_results}), we have also selected the events based on the quality of the PSF, choosing the 3 best partitions (PSF$\,$1 to 3: \texttt{evtype=56}).
To minimise the contamination by Earth limb photons, $\gamma$-ray events with reconstructed directions pointing above a $90\degr$ zenith angle have been excluded.
Standard filters on the data quality were applied.

A \texttt{binned} maximum-likelihood spectral analysis was performed to constrain the high-energy emission of nearby point-like sources and diffuse sky components using the \texttt{NewMinuit} optimization algorithm implemented in \texttt{gtlike}. 
In the modelling\footnote{Source models were built using the \texttt{make3FGLxml.py} tool by T.~Johnson, \href{http://fermi.gsfc.nasa.gov/ssc/data/analysis/user/}{http://fermi.gsfc.nasa.gov/ssc/data/analysis/user/}.} of the region of interest (RoI), we have included the standard templates for the Galactic and isotropic backgrounds\footnote{Namely \texttt{gll\_iem\_v06.fits} and \texttt{iso\_P8R2\_SOURCE\_V6\_v06.txt}, \href{http://fermi.gsfc.nasa.gov/ssc/data/access/lat/}{http://fermi.gsfc.nasa.gov/ssc/data/access/lat/}.} and the source spectral models listed in the 4-year \textit{Fermi} catalogue \citep[3FGL,][]{2015ApJS..218...23A, 2016ApJS..223...26A} within a $25\degr$ radius. 
Normalisations and spectral parameters of the sources lying within $5\degr$ from the RoI centre and displaying a Test Statistic (TS\footnote{$\rm{TS}= 2 \ln (\mathcal{L}_{1} / \mathcal{L}_{0})$, $\mathcal{L}_{1}$ and $\mathcal{L}_{0}$ are the likelihood maxima with or without including the target source into the model.}) above $81$
were left free to vary. 
Otherwise, the normalisations of sources considered as variable ({i.e.,} with a variability index ${\geq} 72.44$ as in the 3FGL) were left free if less than $10\degr$ from the centre of the RoI.

	\subsection{Lightcurve constructions} \label{sec:analysis_lc}

\begin{figure}
\includegraphics[scale=1]{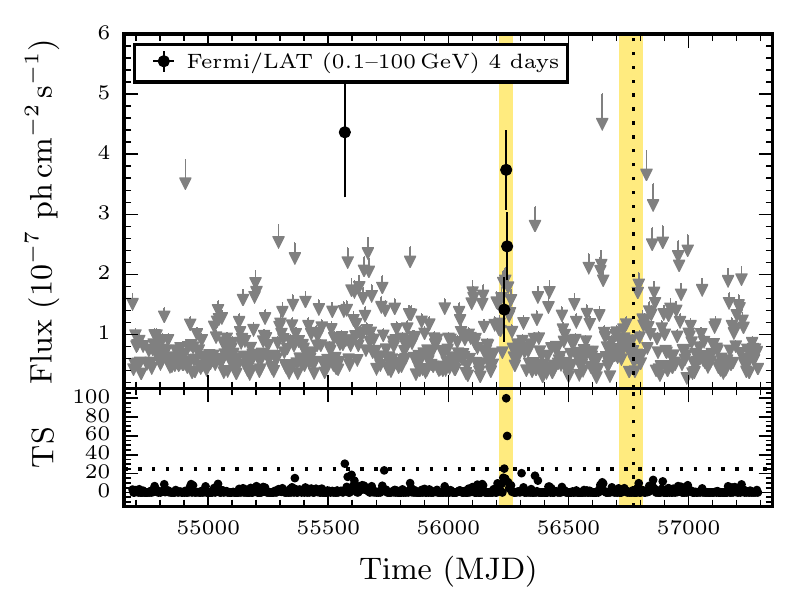}
\vspace{-10pt}
\caption{\hbox{4-day} binned $\gamma$-ray LC at the position of DG$\,$CVn and associated TS values (upper-limit threshold of $\rm{TS}=25$ denoted by the horizontal dotted line). 
Yellow shades represent the time intervals for which more precise LCs were computed (Fig.~\ref{fig:lc_1d_nov12} and \ref{fig:lc_1d_apr14}), while the vertical dotted line marks the superflare occurrence. \label{fig:lc_4d_full}}
\end{figure}	

Lightcurves (LCs) were constructed using the source model derived from the \texttt{binned} maximum likelihood fit performed on the whole \textit{Fermi}/LAT data set (\S\ref{sec:setup}). 
A power-law spectrum point-source model, for which the normalisation and photon index were left free to vary, was added at the position of DG\,CVn.
We performed \texttt{unbinned} maximum likelihood fits on a succession of short time intervals. A $4$-day bin LC was first built over the entire range of available observations to constrain periods when $\gamma$-ray emission can be detected at the localisation of DG$\,$CVn (Fig.~\ref{fig:lc_4d_full}). 
We computed $95$ per cent upper-limits on the high-energy flux (gray arrows) when the TS 
was below $25$ \citep[${\sim} 5 \sigma$,][]{1996ApJ...461..396M} using the (semi-)Bayesian method of \citet{1991NIMPA.300..132H} as implemented in the \texttt{pyLikelihood} module provided with the Science Tools.
Otherwise, integrated $\gamma$-ray fluxes along with $1\sigma$ statistical error bars are provided. 
We estimate the systematic uncertainties to be around $10\%$ in the $0.1$--$100\,$GeV energy range, mainly due to inaccuracies in the effective area characterization.
Over periods of interest, we computed higher precision LCs on $1$-day bins with the same upper-limit computation threshold. 

	\subsection{Test Statistic maps and localisation}\label{sec:tsmap}
The spatial repartition of the high-energy $\gamma$-ray significance level was investigated by calculating TS maps in \texttt{unbinned} mode. 
The significance of an additional point source is evaluated at every position of the $9\degr \times 9\degr$ map with a resolution of $0\fdg1$, with the background source model fixed at the parameters obtained from the global \texttt{binned} analysis.
The position of the $\gamma$-ray excess and the corresponding $68$ per cent statistical confinement radius (${\rm r68}_{\rm stat}$) are determined using the tool \texttt{gtfindsrc}. 
We also report the $95$ per cent confinement circle (r95$_{\rm stat}$) which is computed as $1.6225\,{\rm r68}_{\rm stat}$. 
Following \citet{2015ApJS..218...23A}, we also take into account systematic errors so that ${\rm r68}$ and ${\rm r95}$ are computed as ${\rm r}^2= (1.05\, {\rm r}_{\rm stat})^2 + 0\fdg005^2$.  

\section{Results} \label{sec:results}

The \texttt{binned} likelihood analysis over the full available LAT data set (\S\ref{sec:setup}) easily converged as the RoI lies far away from the Galactic plane (at a latitude of $b=+80\fdg8$).
The normalisation parameters of the diffuse components only diverged by less than $1\%$ from the 3FGL catalogue values.
The goodness of fit was checked by verifying the homogeneity of the residual counts and sigma maps, representing the quantities `$\mbox{model} - \mbox{data}$' and `$(\mbox{model} - \mbox{data})/\sqrt{\mbox{model}}$' respectively. 
Including the DG$\,$CVn source model does not seem essential for the fitting procedure as its derived TS value is ${\sim} 20$. 

Figure \ref{fig:lc_4d_full} shows the LC at the position of DG\,CVn built using \hbox{$4$-day} time bins. 
Four data points exceed the TS threshold of $25$, one around MJD 55570 (\S\ref{sec:firstexcess}) and three around MJD 56240 (\S\ref{sec:nov12_results}). 
There is no significant $\gamma$-ray emission associated with the X-ray/radio superflare of DG\,CVn on 2014 April 23 (MJD 56770.88) (\S\ref{sec:av2014_results}). 
These points are examined in more detail below.

\subsection{January 2011 gamma-ray excess}\label{sec:firstexcess}
The detection at MJD 55570, which reaches a TS value of ${\sim}30$, is time-coincident with a flaring episode of the nearby blazar 3FGL J1332.8$+$2723 reported in the weekly \textit{Fermi All-Sky Variability Analysis} between 2011 January 3 and 10  \citep[FAVA,][]{2013ApJ...771...57A}. 
We found that due to the large PSF of the instrument some of its softest photons spilled over to the position of DG\,CVn, resulting in an artificial TS excess despite the $1\fdg9$ separation with the blazar.

\subsection{November 2012 gamma-ray excess}\label{sec:nov12_results}
\begin{figure}
\includegraphics[scale=1]{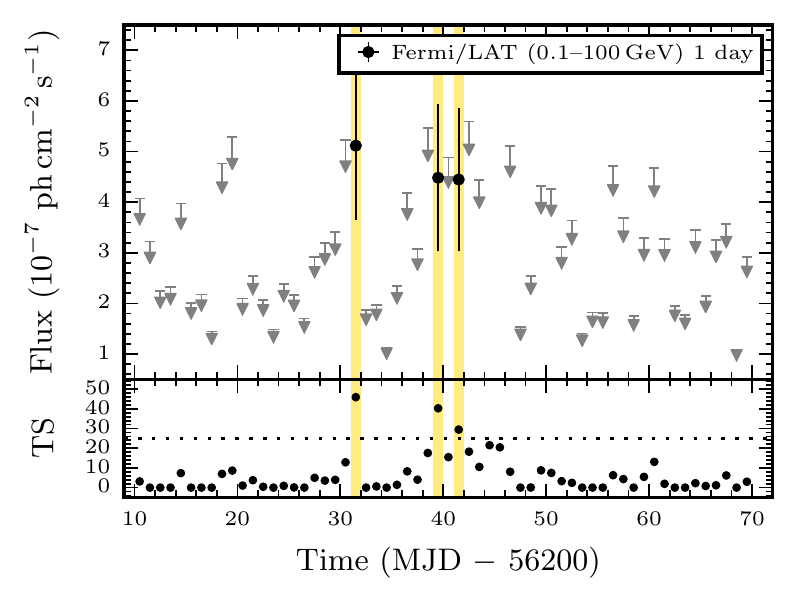}
\vspace{-10pt}
\caption{One-day binned LC built over the $60$-day interval from MJD~$56210$ to $56270$ (1\textsuperscript{st} yellow shaded period comprising the $\gamma$-ray excess in Fig.~\ref{fig:lc_4d_full}). \label{fig:lc_1d_nov12}}
\end{figure}

Three measurements around MJD 56240 (2012 Nov 9) in Fig.~\ref{fig:lc_4d_full} have TS values between $25$ and $100$. 
Again, the FAVA automatic analysis detected a significant transient event for three weeks (from 2012 Oct 29 to Nov 19) and attributed it, as the January 2011 flare, to the blazar 3FGL J1332.8$+$2723. 
As detailed below, we consider this association incorrect.
We also note that this event is not included in the dataset used to build the 3FGL catalogue. Therefore there is no \textit{Fermi}/LAT counterpart despite the large TS value.

We constructed a $1$-day binned LC over $60$ days starting from MJD 56210. 
The selected time-scale is represented by a yellow shaded vertical band in Fig.~\ref{fig:lc_4d_full}. 
The resulting LC shows the $\gamma$-ray flare evolved over several days (Fig.~\ref{fig:lc_1d_nov12}). 
The addition of a point-source at the position of DG\,CVn is significant with a daily TS value up to $46$. 
It starts on MJD 56231 with a peak flux of $(5.1 \pm 1.5) \times 10^{-7}\,$ph$\,$cm$^{-2}\,$s$^{-1}$ and then quenches for a week before a re-brightening on MJD 56238 at a similar flux level. 

\begin{figure}
\includegraphics[scale=1]{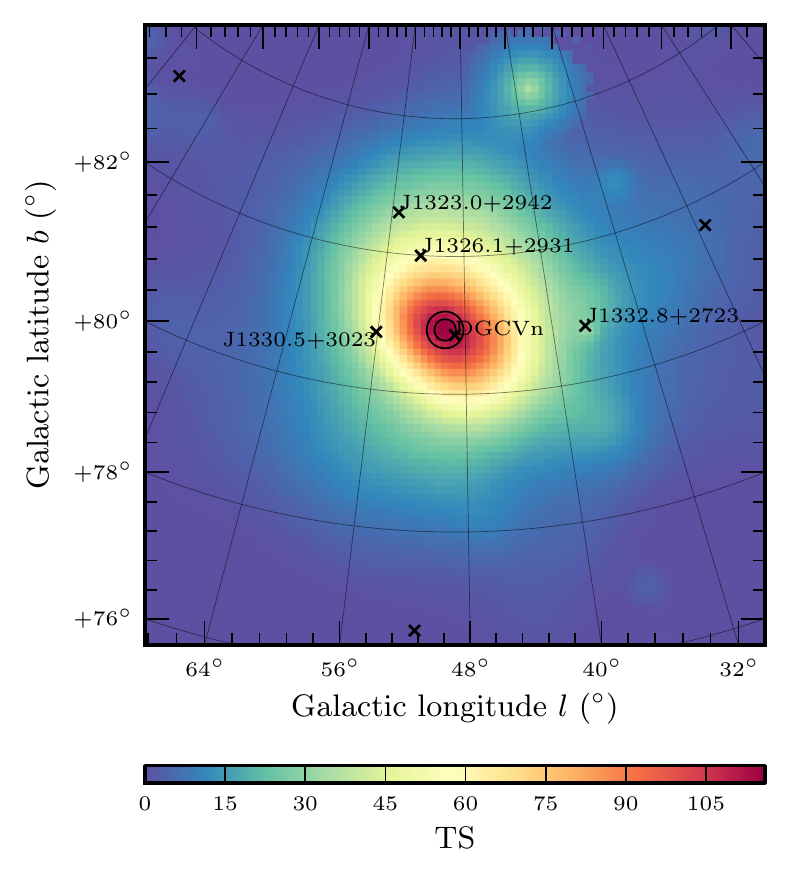}
\vspace{-10pt}
\caption{$8\degr \times 8\degr$ residual TS map ($0\fdg1\,$pixel$^{-1}$) computed on the time interval highlighted in Fig.~\ref{fig:lc_1d_nov12}. $68$ and $95$ confidence radii on the best-fit position are overplotted. \label{fig:tsmap_stacked}}
\end{figure}

To precisely locate the emission origin, we created residual TS maps (\S\ref{sec:tsmap}) for $20$ individual days encompassing the flare around MJD 56240. 
To increase the sensitivity, we stacked together the data corresponding to the three days when DG\,CVn's model addition yields a TS peak above $25$ ({i.e.,} 2012 Oct 31, Nov 8, 10, all yellow-shaded in Fig.~\ref{fig:lc_1d_nov12}). 
An \texttt{unbinned} likelihood analysis was then performed along with a residual TS map computation (see Fig.~\ref{fig:tsmap_stacked}). 
The TS map shows that a source is detected with a highly-significant TS of $116$ at a best fit position of $\mbox{RA}=202\fdg83$, $\mbox{Dec.}=29\fdg41$, with containment radii $\mbox{r68}=0\fdg16$ and $\mbox{r95}=0\fdg26$. 
The source spectrum is a power-law photon index of $2.37 \pm 0.18$ for a mean $\gamma$-ray flux of $(4.6 \pm 0.8) \times 10^{-7}\,$ph$\,$cm$^{-2}\,$s$^{-1}$.

The location of the source clearly excludes an association with the blazar 3FGL J1332.8$+$2723 ($2\degr$ away from the best-fit position, Fig.~\ref{fig:tsmap_stacked}). 
It is also distinct from the closest known sources (namely 3FGL J1326.1$+$2931 and 3FGL J1330.5$+$3023, a.k.a. 3C$\,$286), although we note that, because a significant portion of the flare photons encroaches upon these objects, a $\gamma$-ray excess around MJD$\,56240$ is visible in their public $30$-day LCs\footnote{Aperture photometry LCs of 3FGL sources with $30\,$day time resolution are weekly updated and available on the \textit{Fermi} Science Support Center (\texttt{http://fermi.gsfc.nasa.gov/ssc}).}. 
However, an association with DG\,CVn remains possible since the binary is $0\fdg17$ away from the best-fit position, just outside the $68\%$ confidence region (Fig.~\ref{fig:tsmap_stacked}).


	\subsection{April 2014 superflare counterpart}\label{sec:av2014_results}
\begin{figure}
\includegraphics[scale=1]{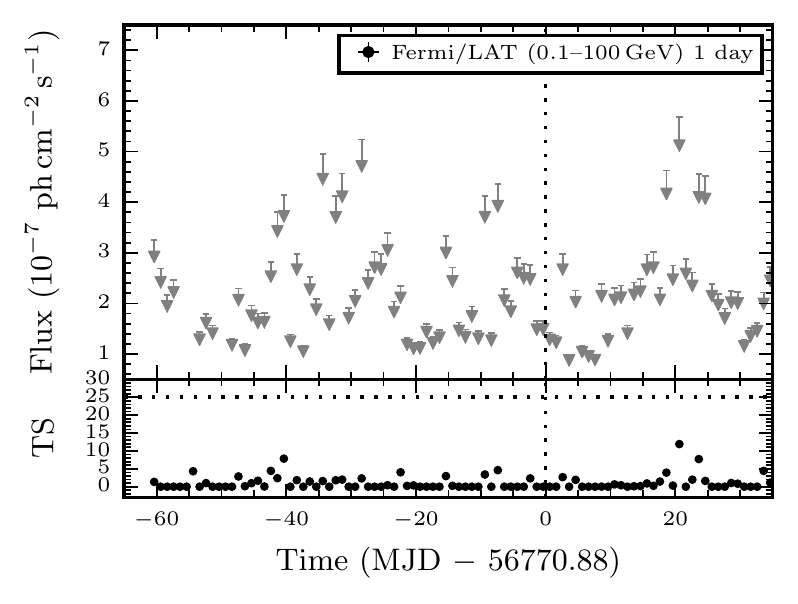}
\vspace{-10pt}
\caption{One-day binned LC built over the $100\,$days interval (2\textsuperscript{nd} yellow shaded period in Fig.~\ref{fig:lc_4d_full} encompassing the X-ray superflare (the time reference is set to the \textit{Swift} trigger date: 2014 Apr 23). \label{fig:lc_1d_apr14}}
\end{figure}

We constructed the LC of DG$\,$CVn on \hbox{$1$-day} bins, starting 70 days prior to the X-ray/radio superflare occurrence (MJD 56770.88) and ending 30 days after to cover possible delays (Fig.~\ref{fig:lc_1d_apr14}). 
The most significant measurement occurs twenty days after the X-ray flaring episode and has a TS value around $12$ with a statistical fluctuation probability ${\sim} 22\%$ (assuming 100 independent trials). 
This measurement corresponds to an upper limit on the $\gamma$-ray flux of $5.7 \times 10^{-7}\,$ph$\,$cm$^{-2}\,$s$^{-1}$. 
We conclude that there is no significant $\gamma$-ray emission at the location of DG$\,$CVn in April 2014.

\section{Discussion} \label{sec:discussion}
We did not find evidence for $\gamma$-ray emission from DG\,CVn during its superflare in April 2014, but we detected significant flaring emission in November 2012 from a direction compatible with the location of DG\,CVn. We now discuss the possible origin of this emission.

	\subsection{Association with DG$\,$CVn?}\label{sec:association}
One possible interpretation of the flaring episode around MJD$\,56240$ is a series of stellar eruptions associated with DG\,CVn, each of them lasting for less than a day. 
On the one hand, active stars are not known to produce such high-energy and long-lasting outbursts, and no concurrent flaring has been reported at any other wavelength. 
On the other hand, the April 2014 superflare \citep{2014ATel.6121....1D} as well as the radio emission \citep{2015MNRAS.446L..66F} were not expected. 
A major outburst might have happened and remained unnoticed by a lack of coincident monitoring. 

Due to the proximity of our Sun, the LAT is able to detect solar flares above tens of MeV \citep[see for e.g.][for a list of several detected flares and a study of long lasting emissions]{2014ApJ...789...20A, 2014ApJ...787...15A}. The impulsive/prompt phase is easier to detect at high-energy \citep[with a transient event lasting for a few minutes to less than an hour,][]{2011ATel.3552....1O} because of the increased flux level, but sometimes the emission can extend to several hours, as presented in \citet{2012ATel.3886....1T}. 
The solar flare with the highest $\gamma$-ray flux in the LAT reached ${\sim} 4 \times 10^{-7}\rm\,erg\,cm^{-2}\,s^{-1}$ for a typical X-ray flux (X class) of ${\sim} 0.1 \rm\,erg\,cm^{-2}\,s^{-1}$ \citep{2014ApJ...787...15A}. 
The November 2012 flare reached a ${>}100\,$MeV luminosity of ${\sim} 10^{31}\rm\,erg\,s^{-1}$ (at $18$\,pc).
Assuming the same flux ratio as this solar flare, any accompanying X-ray flare would have been spectacular and unlikely to be missed by all-sky monitors (a quicklook daily analysis with \textit{Fermi}/GBM does not reveal any contemporaneous bright hard X-ray emission in the $12$--$100$\,keV band). 
Inversely, the predicted $\gamma$-ray flux is orders of magnitude too small (${\sim} 10^{-10}$\,ph\,cm$^{-2}$\,s$^{-1}$) to be detectable when scaling to the peak X-ray flux of ${\sim} 3 \times 10^{-9}\rm\,erg\,cm^{-2}\,s^{-1}$ observed during the April 2014 flare of DG\,CVn  (corresponding to an X-ray luminosity ${\sim} 10^{32}\rm\,erg\,s^{-1}$, \citealt{2015MNRAS.446L..66F}). 
If the November 2012 excess is associated with flaring activity in DG\,CVn then the mechanism is entirely different from that at work in solar flares, with extremely efficient conversion of flare energy into $\gamma$-ray emission.

Such a mechanism is all the more unlikely that we have also looked for similar high-energy flaring behaviour among a selection of active binary stars, all classified as RS$\,$CVn or Algol variables (namely EV$\,$Lac, UX$\,$Ari, HR$\,$1099, Algol, II$\,$Peg, HR$\,$5110, V374$\,$Peg, GJ$\,$2036A, V857$\,$Cen, GL$\,$Vir, HU$\,$Del, GJ$\,$3225, GJ$\,$3153, G$\,$180$-$11 and EQ$\,$Peg). 
They were chosen based on their proximity (the farthest one lies at $50\,$pc), their remoteness from the Galactic plane ($|b|>13\degr$) and rapid rotation as a proxy for chromospheric activity ($v \sin i$ above tens of km$\,$s$^{-1}$). 
No evidence for $\gamma$-ray flares was found over seven years of the \textit{Fermi} mission. 

We conclude, based on the multi-wavelength picture, flare energetics, and lack of comparable behaviour in other systems, that the November 2012 $\gamma$-ray flare is very unlikely to be associated with DG$\,$CVn.

	\subsection{AGN flaring event}
The November 2012 flare may have been caused by a background AGN. 
AGNs make up more than $71\%$ of high-latitude ($|b| {>} 10\degr$) \textit{Fermi}/LAT sources \citep{2015ApJ...810...14A}. 
Among them, $98\%$ are blazars (either Flat Spectrum Radio Quasars FSRQs or BL Lacertae objects). 
Blazar LCs are known for their variability on a wide range of time-scales, with the strong flaring interpreted as internal shocks and/or sporadic changes in the physical conditions of the relativistic jet. 
For instance, the blazar 3C~279 underwent multiple distinct flares in 2013--2014 observed in $\gamma$ rays \citep{2015ApJ...807...79H} with significant variability observed over a few hours. 
The derived power-law photon indices range from $1.71\pm0.10$ to $2.36\pm0.13$. 
The properties of the transient excess we found (\S\ref{sec:nov12_results}) could be compatible with these characteristics, although we caution that our uncertainties prevent any clear classification. 

The probability to find a $\gamma$-ray AGN in the vicinity of DG CVn depends on their Log $N$ -- Log $S$ distribution. 
As a rough estimate, we note that the integrated TS reaches ${\sim}20$ (\S\ref{sec:results}) so that the flaring source is close to being included in the \textit{Fermi}/LAT catalogue. 
Assuming that the 3FGL catalogue is complete at latitudes $|b| {>} 10\degr$ and that the 2193 sources that are listed at such latitudes are AGNs, we expect $1.4 \times 10^{-2}$ background blazars within the $0.21\,$deg$^2$ solid angle corresponding to the $2\sigma$ confidence region. 
This is small but not statistically implausible. 

We have searched for AGN counterparts within the \textit{Veron Catalogue of Quasars \& AGN, 13\textsuperscript{th} Edition} \citep{2010A&A...518A..10V}, the \textit{5\textsuperscript{th} Edition of the Roma BZCAT Multi-frequency Catalogue of Blazars} \citep{2009A&A...495..691M}, the \textit{WISE Blazar-like Radio-Loud Source (WIBRaLS)} catalogue \citep{2014ApJS..215...14D} and the \textit{CRATES Flat-Spectrum Radio Source Catalogue} \citep{2007ApJS..171...61H}. 
None of the few matches in the Veron catalogue correspond to X-ray sources in the \textit{3XMM-DR5} catalogue \citep{2016yCat.9046....0X} or to radio sources in the \textit{FIRST} survey catalogue \citep{2015ApJ...801...26H}. 
Hence, there is no obvious candidate counterpart amongst catalogued AGNs.

We then searched for radio counterparts in the error circle of the $\gamma$-ray source to identify possible blazar candidates. 
The \textit{FIRST} survey returns 19 sources, five of which having $1.4$\,GHz fluxes above $10$\,mJy. 
Where available, we investigated their radio spectrum using \textit{SPECFIND} \citep{2010A&A...511A..53V}. Most sources are faint and lack multi-frequency observations but one source, FIRST J133101.8$+$293216, has a radio spectrum indicative of a blazar, with a spectral index $\alpha\ga -0.5$ (defined as $F_\nu \sim \nu^\alpha$). 
Indeed, the source is selected in the sample of FSRQs assembled by \citet{2003ApJ...594..684M}. 
The source has a flux density of 136$\pm$27 mJy at $325$\,MHz, $43.8\pm8.8$\,mJy (FIRST) or $35.2\pm7.0$\,mJy (NVSS) at $1.4$\,GHz, $24.6\pm4.9$\,mJy at $4.85$\,GHz. 
This radio source has a matching \textit{SDSS} source \citep[SDSS J133101.83$+$293216.5 in DR12 with a photometric redshift $z=0.48$,][]{2015ApJS..219...12A} and a matching source in the \textit{AllWISE} catalogue \citep[WISE J133101.82$+$293216.3,][]{2014yCat.2328....0C}. 
The IR colours are close to those of the \textit{Fermi}-detected FSRQ PMN J2023$-$1140 \citep{2012ApJ...748...68D}.
There is an X-ray source, 1WGA~J1331.1$+$2930, in the WGA ROSAT catalogue located $1\farcm3$ away from this radio/optical/IR source, with a quoted position uncertainty of $50\arcsec$ \citep{2000yCat.9031....0W}. The X-ray flux is $3.4\times 10^{-13}\rm\,erg\,cm^{-2}\,s^{-1}$ based on $7$\,ks of exposure. We found no other information on this X-ray source. Taken at face value, the radio and X-ray fluxes -- if associated and representative of the average fluxes -- are consistent with a low-luminosity FSRQ \citep{2011ApJ...743..171A}.
Given the currently available multiwavelength data, this radio source situated $0\fdg14$ away from the \textit{Fermi}/LAT localisation, within the $68\%$ confidence region, is a plausible candidate counterpart to the November 2012 $\gamma$-ray flare.

\section{Conclusion}
Motivated by the energetic stellar flare detected in radio and X-rays from DG\,CVn, we have searched for $\gamma$-ray emission at the location of this system over seven years of {\it Fermi}/LAT operations. 
There is no $\gamma$-ray emission associated with the April 2014 superflare of DG\,CVn. 
$\gamma$-ray emission is detected in January 2011 but is attributed to a nearby flaring blazar, 3FGL J1332.8$+$2723. 
Flaring emission is also detected in November 2012 with a location consistent with that of DG\,CVn. 
However, the lack of reported simultaneous flaring at other wavelengths from DG\,CVn, together with general considerations on the energetics of stellar flares, make it unlikely that this $\gamma$-ray emission originated from this system. 
Inspection of catalogues reveals a more mundane explanation for the November 2012 flare in the form of a plausible blazar candidate within the $\gamma$-ray error circle. 
Additional observations in radio, optical and X-rays and/or of additional $\gamma$-ray activity will be required to establish the spectrum, variability and redshift of this source and secure its identification as a blazar.

\section*{Acknowledgments}
We thank the anonymous referee for his/her thorough review and for pointing out the X-ray source 1WGA J1331.1+2930 to us.
AL and SC acknowledge funding support from the French Research National Agency: CHAOS project ANR-12-BS05-0009 and the UnivEarthS Labex program of Sorbonne Paris Cit\'e (ANR-10-LABX-0023 and ANR-11-IDEX-0005-02). 
AL thanks P.~Jenke, V.~Connaughton and C.~Wilson-Hodge for the analysis of \textit{Fermi}/GBM data.
GD thanks X.~Delfosse for useful discussions concerning DG\,CVn. This research has made use of the VizieR catalogue access tool, CDS, Strasbourg, France. The original description of the VizieR service was published in A\&AS 143, 23.
The \textit{Fermi} LAT Collaboration acknowledges generous ongoing support from a number of agencies and institutes that have supported both the development and the operation of the LAT as well as scientific data analysis.
These include the National Aeronautics and Space Administration and the Department of Energy in the United States, the Commissariat \`a l'Energie Atomique and the Centre National de la Recherche Scientifique / Institut National de Physique Nucl\'eaire et de Physique des Particules in France, the Agenzia Spaziale Italiana and the Istituto Nazionale di Fisica Nucleare in Italy, the Ministry of Education, Culture, Sports, Science and Technology (MEXT), High Energy Accelerator Research Organization (KEK) and Japan Aerospace Exploration Agency (JAXA) in Japan, and the K.~A.~Wallenberg Foundation, the Swedish Research Council and the
Swedish National Space Board in Sweden.
Additional support for science analysis during the operations phase is gratefully acknowledged from the Istituto Nazionale di Astrofisica in Italy and the Centre National d'\'Etudes Spatiales in France.

\bibliographystyle{mnras}
\bibliography{DGCVn_Paper}

\bsp 

\label{lastpage}

\end{document}